\documentclass[aps,preprint,superscriptaddress,prx,longbibliography]{revtex4-2}
\usepackage{graphicx,amsmath,subcaption}

\usepackage{amsmath}
\usepackage{epsf}
\usepackage{graphicx}

\usepackage{dcolumn}
\usepackage{bm}
\usepackage{subcaption}
\usepackage{amssymb}
\usepackage{array}
\usepackage{varioref}
\usepackage{color}
\usepackage[dvipsnames]{xcolor}

\usepackage[colorlinks, linkcolor=blue, citecolor=red]{hyperref}
\usepackage{soul}
\usepackage{float}
\restylefloat{table}

\usepackage{babel}
\usepackage{natbib}

\providecommand{\pd}{\partial}

\providecommand{\bv}[1]{\bm{\mathrm{#1}}}

\providecommand{\ke}{\epsilon}

\begin{document}

\title{
%Geometrically induced multiferroic properties in plastically deformed SrTiO$_3$ \IS{alternative title 
Multiferroicity in plastically deformed SrTiO$_3$}

\author{Xi Wang}
\thanks{Equal contribution}
\affiliation{Department of Physics, Bar-Ilan University, 52900, Ramat Gan, Israel}
\affiliation{Institute of Nanotechnology \& Advanced Materials, Bar-Ilan University, 52900, Ramat Gan, Israel}
\author{Anirban Kundu}
\thanks{Equal contribution}
\affiliation{Physics Department, Ariel University, Ariel 40700, Israel}
\author{Bochao Xu}
\thanks{Equal contribution}
\affiliation{Department of Physics, University of Connecticut, Storrs, Connecticut 06269, USA}
\author{Sajna Hameed}
\thanks{Present address: Max Planck Institute for Solid State Research,
Heisenbergstraße 1, 70569 Stuttgart, Germany}
\affiliation{School of Physics and Astronomy, University of Minnesota, Minneapolis, MN, 55455, USA}
\author{Ilya Sochnikov}
\affiliation{Department of Physics, University of Connecticut, Storrs, Connecticut 06269, USA}
\affiliation{Institute of Materials Science, University of Connecticut, Storrs, Connecticut, USA}
\affiliation{Materials Science and Engineering Department, University of Connecticut, Storrs, Connecticut, USA}
\author{Damjan Pelc}
\affiliation{Department of Physics, Faculty of Science, University of Zagreb, Zagreb, HR-10000, Croatia}
\author{Martin Greven}
\affiliation{School of Physics and Astronomy, University of Minnesota, Minneapolis, MN, 55455, USA}
\author{Avraham Klein}
\thanks{Corresponding author}
\affiliation{Physics Department, Ariel University, Ariel 40700, Israel}
\author{Beena Kalisky}
\thanks{Corresponding author}
\affiliation{Department of Physics, Bar-Ilan University, 52900, Ramat Gan, Israel}
\affiliation{Institute of Nanotechnology \& Advanced Materials, Bar-Ilan University, 52900, Ramat Gan, Israel}

%\maketitle

%\clearpage

\begin{abstract}
%\IS{(The abstract needs work, it does not completely follow the classic structure Background-Problem-Method-Findings-Implications and it would be good to make it more exciting for the glossy journals. Can try editing after the main text is more or less finalized.)} 
A major challenge in the development of quantum technologies is to induce additional types of ferroic orders 
%magnetic properties 
into materials that exhibit other useful quantum properties. Various techniques have been applied to this end, such as elastically straining, doping, or interfacing a compound with other materials.
%, or straining a parent e.g. by doping a parent quantum material or interfacing it with another material.
%like ferroelectricity.https://www.overleaf.com/project/636d43a41fb5671594594a4c
%into preexisting states, such as superconductivity \IS{(Mentioning of SC feels disconnected from the rest, I know why it is mentioned, but it is worth trying a different opening sentence as we don't show any SC data.)}. 
%Strain engineering is a well-known avenue to this end, but the application of reversible, elastic strain has only yielded limited results. 
Plastic deformation introduces permanent
%, whereby 
topological defects and large local strains 
%are introduced 
into a material, which 
%potentially 
can give rise to qualitatively new functionality.
Here we show via local magnetic imaging that plastic deformation induces robust magnetism in the quantum paraelectric SrTiO$_3$, in both conducting and insulating samples. 
%Our results are verified by two independent groups performing local magnetic imaging with independently grown crystals and a variety of doping methods.
Our analysis indicates that the magnetic order is localized along dislocation walls and coexists with polar order along the walls. 
%Both orders are intrinsic and induced by the strong geometric deformations causing the dislocations. The mutual interaction of the two induces a helical order for the magnetism, and the responsiveness of the polar order to elastic strains provides a method of controlling its moment, allowing its out-of-plane component to be switched on and off.
The magnetic signals can be switched on and off in a controllable manner with external stress, which demonstrates that plastically deformed SrTiO$_3$ is a quantum multiferroic. 
These results establish plastic deformation as a versatile platform for quantum materials engineering.

\end{abstract}
% \keywords{Suggested keywords}%Use showkeys class option if keyword
% display desired
\maketitle

One of the most sought-after properties of quantum materials is the manifestation of multiple order parameters, either in coexistence or by ``switching'' between states via external stimuli. Since charge, spin, superconducting, topological or other orders are crucial for quantum technologies, multi-order systems are expected to be of significant value to future applications. 
One pathway to creating multi-order systems is to take a compound with known properties and introduce additional orders via doping or interfacing with a different compound \cite{Giustino2020,Hwang2012,Zunger2021}. %\cite{Li2011,Tan2013}{\bf references needed here; do we really need 16 reference in the next two sentences?}. 
An alternative route that has
been receiving much recent attention is strain engineering.
%\cite{Zubko2007, Ahadi2019, Schlom2007,Herrera_2019,Xu2020}
%\ani{[1,3,5,7(or 4),8, if this is also long suggest to remove 8 ]}\cite{Zubko2007, Chen2021,Ahadi2019, franklin2020giant, Schlom2007,Xue2016,Herrera_2019,Xu2020}. 
It has been used, or proposed as an avenue, to induce magnetism, ferroelectricity and other ordered states, to enhance superconductivity, and to create novel excitations \cite{Maiti2003, Schlom2007,  MacManus-Driscoll2008, Tan2013, Ahadi2019, Herrera_2019,Nova2019,   Liu2021}.
%\ani{[10,13,15]}\cite{Maiti2003,MacManus-Driscoll2008,Tan2013,Liu2021,Nova2019,Khaetskii_2021,PhysRevLett.122.057208,basini2022}.
However, this approach typically involves relatively small elastic strains
and is not a permanent modification -- once the stress is removed, the new properties vanish.
In contrast, plastic deformation leads to extremely large local strains and induces permanent modifications. This approach is widely used in materials engineering \cite{Hansen2014,Kumar2021b,Sypek2017,
Xiao2021}, but has received comparatively little attention in the quantum materials community.
%as the 
Conventional wisdom holds that quantum materials, especially metallic and superconducting systems, should be as clean and pristine as possible, while plastic deformation induces many defects and strong geometric deformations. However, a recent study found that, in stark contrast to this preconception, plastic engineering can induce a permanent enhancement of superconductivity and ferroelectricity in the 
unconventional superconductor SrTiO$_3$ (STO) \cite{Hameed2022}. This intriguing discovery suggests that the plastic engineering avenue 
needs to be far more thoroughly explored \cite{Li2017,Li2019}.

STO is a well-known unconventional superconductor and a quantum-critical incipient ferroelectric (FE) \cite{Scheerer2020} that maintains its inversion-breaking properties upon doping with carriers \cite{Zubko2007, Song2019, Xu2020,enderlein2020superconductivity,Gastiasoro2020}. Pristine STO is typically considered nonmagnetic and exhibits a weak bulk diamagnetic response. However, there have been persistent hints of emergent magnetism in STO and various STO-based interfaces
\cite{Haeni2004, Zubko2007,Pai2018,LI2011,Bert2011,kalisky2012,
Christensen2019,basini2022, Coey_2016,Bi2014,Park2020}.
In addition, STO crystals are surprisingly ductile even at ambient temperature, with plastic deformation levels up to $\sim10$\% achieved
in both insulating and metallic samples \cite{Gumbsch2001, Hameed2022}. Plastic deformation leads to the formation of self-organized dislocation structures -- dislocation walls -- surrounded by a field of high strain, where local atomic displacements are comparable to the unit cell dimensions \cite{Chen2021,Hameed2022}. This material is therefore a prime candidate to obtain emergent properties through plastic strain engineering.  

Here we show that plastic deformation generates intrinsic magnetism in STO. We perform magnetic measurements on the micron scale and find 
a robust stripe-like magnetic landscape in both insulating and conducting STO. We attribute this pattern to the localization of magnetism at the dislocation walls that run perpendicular to the deformation axis. 
By comparing our data with previous measurements of local ferroelectricity near the dislocations \cite{Hameed2022}, we demonstrate that both ferroelectric and magnetic orders are intrinsic and arise from magneto- and polar- elastic interactions.
Furthermore, our theoretical analysis shows that the two orders are strongly correlated and thereby establishes that plastically deformed STO is a quantum multiferroic.

\begin{figure*}
\centering
\includegraphics[width=1\textwidth]{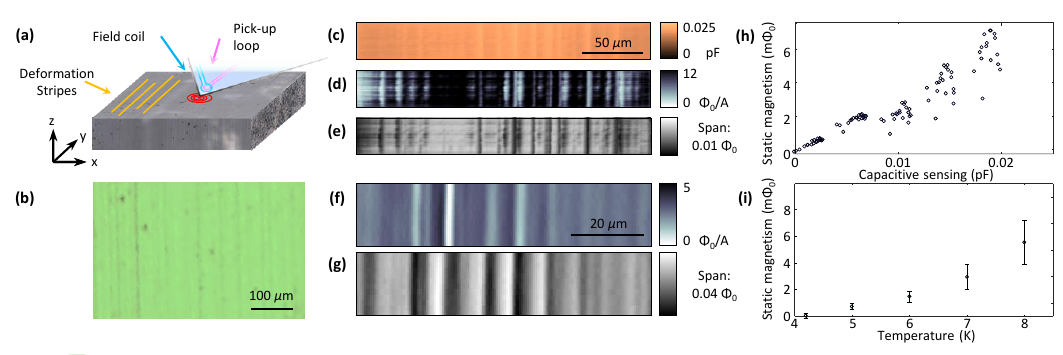}
\caption{Scanning SQUID measurement and stripy magnetic order in plastically deformed SrTiO${_3}$ observed by scanning in contact. (a) Illustration of scanning SQUID measurement in contact mode  \cite{note1}. (b) Optical images of the sample. The deformation stripes are visible on the surface. (c-e)  Scanning SQUID images of capacitive surface sensing, susceptibility, spontaneous magnetism, respectively, in an undoped sample. The magnetic stripe direction coincides with that of the deformed stripes shown in panel b. (f-g) Susceptibility and spontaneous magnetism, respectively, in a conducting sample (SrTi${_{1-x}}$Nb${_x}$O${_3}$, $x=0.004$). (h) Stress dependence of the magnetism exhibits linear behaviour (see also Fig. \ref{fig:exp_dependence}a). (i) The strength of the magnetism (see also \ref{fig:exp_dependence}b) increases with increasing temperature.\label{exp_illustration}}
\end{figure*}

\section*{Stripe magnetism in plastically deformed STO}

In this study, we examine three types of plastically deformed STO: metallic samples doped with niobium or oxygen vacancies (OVD), and undoped, insulating samples. In order to map the local magnetic landscape of deformed STO we employ scanning superconducting quantum interference device (SQUID) microscopy between 5 K and 16 K, a temperature range in which quantum FE fluctuations are expected to be strong \cite{rowley2014ferroelectric,Christensen2019}.
The sketch in Fig. \ref{exp_illustration}a illustrates our measurement setup (see Methods for more details). The scanning SQUID contains a micron-sized pick-up loop on a tip and a coil to generate external magnetic fields. This enables the detection of both spontaneous and field-generated magnetism with a high spatial resolution.
%of \bf???\rm. AK: micron scale (from above)
The tip-to-sample distance can be varied, including to a zero-distance configuration where the tip is in mechanical contact with the sample.
This approach allows us to investigate and characterize the magnetic properties of plastically deformed STO at the micron scale.

%\vi{(note: discussion on Fig 1s needs to be reordered)}

\begin{figure*}
\centering
\includegraphics[width=1\textwidth]{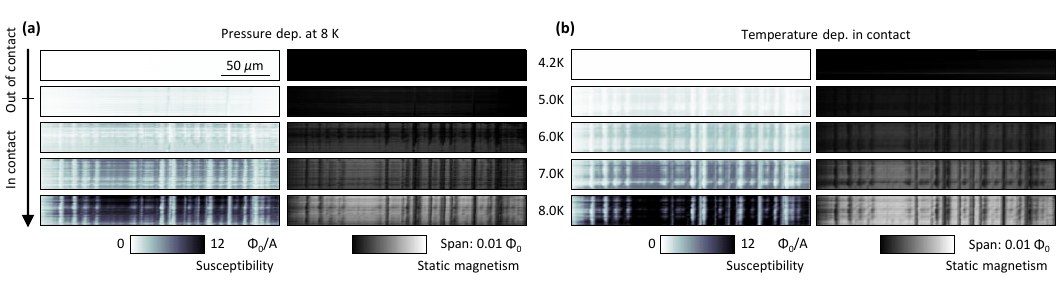}
\caption{Stress and temperature dependence of magnetic signal in undoped, plastically deformed SrTiO${_3}$. (a) Stress dependence of magnetic stripes at 8 K. Before contact with the surface by the SQUID tip, we observe only background noise in the measurement. The magnetic response increases with the applied force. (b) Temperature dependence of magnetic stripes in susceptibility and static magnetism by scanning in contact. The applied force is estimated to be 0.1 - 1 $\mu$N. (Note: this range is rather uncertain because of the nonuniform surface.) The magnetic response increases with the temperature.\label{fig:exp_dependence} }
\end{figure*}

We report on two main findings. First, we detect robust stripe-like signatures of magnetism, but \emph{only} when the tip is in contact with the sample. This is seen in  
Figs. \ref{exp_illustration}d-e, which show SQUID images of both magnetic order and paramagnetic
%and 
susceptibility as a function of two-dimensional space when scanning in contact, with similar results for both observables. The stripe-like modulation is also found in the stress-sensing capacitance channel, as shown in Fig. \ref{exp_illustration}(c). The orientation of stripes in the SQUID images is consistent 
with the direction of dislocation walls formed during deformation (%optical picture is in 
see Fig. \ref{exp_illustration}b and discussion below). 
%On the contrary, when scanning in non-contact mode, no magnetism is observed on the surface, 
Within instrumental resolution, no magnetic signal is found in non-contact mode (Fig. \ref{fig:exp_dependence}a).
%and the SQUID images show the SQUID noise. 
The magnetic signal increases linearly with applied stress
%as the applied stress is increased, the magnetic signal strength shows a linear dependence on stress 
(Fig. \ref{exp_illustration}h), accompanied by enhanced contrast in the magnetic stripes (Fig. \ref{fig:exp_dependence}a), before saturating at an 
%However, at the regime of higher applied force, the magnetic response to stress saturates near an 
applied force of approximately  $1\mu$N.
%and ceases to increase. 
%This saturation could originate from a saturation of the applied stress as a consequence of bending the soft cantilever. \BK{last sentence belongs to supp if we have one}

Our second main finding is that the in-contact magnetic signal strength increases with temperature up to at least 16 K (Fig. \ref{fig:exp_dependence}b and SI). 
This is in contrast to the conventional situation whereby magnetic order decreases with increasing temperature. 
%\IS{We note that the SQUID is influenced by the sample upon contact at these elevated temperatures and therefore the quantitative temperature dependence of the magnetic signals should be only looked at at lower temperatures below about 8 K. However, we do believe that the magnetism persists likely well above 16 K.}  
% AK: IS - this line can go in the SI
%\IS{(Comment: I think this was mentioned in the previous iterations that DM scenario predicts enhancement of magnetism with T, we should probably mention that here or somewhere in the discussion and cite relevant works)}
%We note that similar $T$ dependence, albeit with much weaker signal, was previously reported in both STO and LAO/STO \cite{Christensen2019}.
%is an unexpected result considering that paramagnetism typically decreases with rising temperature due to the interplay between self-energy and thermal fluctuation. \BK{[Avi, can we say the previous sentence?]}
%\st{Interestingly, the increasing signal  with temperature is found near 4 K, well above the bulk superconducting transition.}
In our present setups, it is not possible to perform measurements at higher sample temperatures, at which the SQUID loop cannot be kept superconducting while the cantilever is in physical contact with the sample. 
The observed behaviour is consistent across all our measured samples - both insulating and (super)conducting. We do note that, in one Nb-doped sample,
%(UMN), 
we found magnetism
%the 
%magnetism was only found 
only in a small region on the crystal surface (see SI). 

These two principal observations are highly robust, since the same results are obtained by two independent scanning SQUID groups, in different samples from distinct sample sources. The magnetic signal is roughly two orders of magnitude stronger than in previous measurements of in-contact magnetic signal in
%upon contact in 
LaAlO$_3$/STO interfaces and pristine STO domain walls
%, which is modulated over the tetragonal domain structure 
\cite{Christensen2019}. We note that an unconventional signal increase with temperature was found in those experiments as well. If we assume that the observed magnetic signal stems from some effective surface magnetization, $m_S$, we estimate $m_S \sim1~\mu_B/a_0^2$, where $\mu_B$ is the Bohr magneton and $a_0 \approx 4 \mathring{\rm A}$ is the STO lattice constant (see Methods). This conservative estimate is not based on any specific theoretical microscopic consideration.
%per unit cell area
%\BK{Xi estimates 0.5, still working on it} XX $\mu_B$ per surface \IS{(dislocation plain??? is the unit cell of the dislocation is larger??)} unit cell \IS{(If it is 0.5 mub, that's pretty large for a nominally non-magnetic material, would be an important thing to mention in the abstract, I'd say that is a strong magnetism, if the estimate is true - seems like a big deal)}.
%in deformed STO, XX times higher than previous observation in LAO/STO sample.

In order to understand the observed phenomena, we now briefly mention some pertinent and previously reported information on some of the same samples as those studied here \cite{Hameed2022}. First, neutron and x-ray diffuse scattering has established the existence of dislocation walls running perpendicular to the deformation axis.
%(\bf why not use same coordinate system as in our 2022 paper: stress along [010]?! \rm along [100]). 
% AK: because of the way we did the theoretical analysis. Maybe we can change this around at a review stage.
The walls consist of pairs of dislocations with Burgers vectors along the unit cell face diagonal,
%(\bf change this ? \rm 
(e.g. [101], [10-1]), with an inter-dislocation distance of approximately $h = 14a_0$ \cite{note1}.
%, where $a_0$ is the lattice constant. 
Thus the unit cell of a wall has a length of $2h$. Second, Raman scattering has established the existence of static polar order, believed to be localized near the walls. The polar order increases monotonically with decreasing temperature down to 7 K, the base temperature of that experiment. In addition, low-frequency Raman scattering has shown that these samples also host bulk quantum critical fluctuations that coexist with the localized order.

\section*{Coupling of dislocations, magnetism, and ferroelectricity}

\begin{figure*}
\centering
\begin{minipage}{0.29\hsize}
\begin{subfigure}{\hsize}
        \includegraphics[width=\hsize]{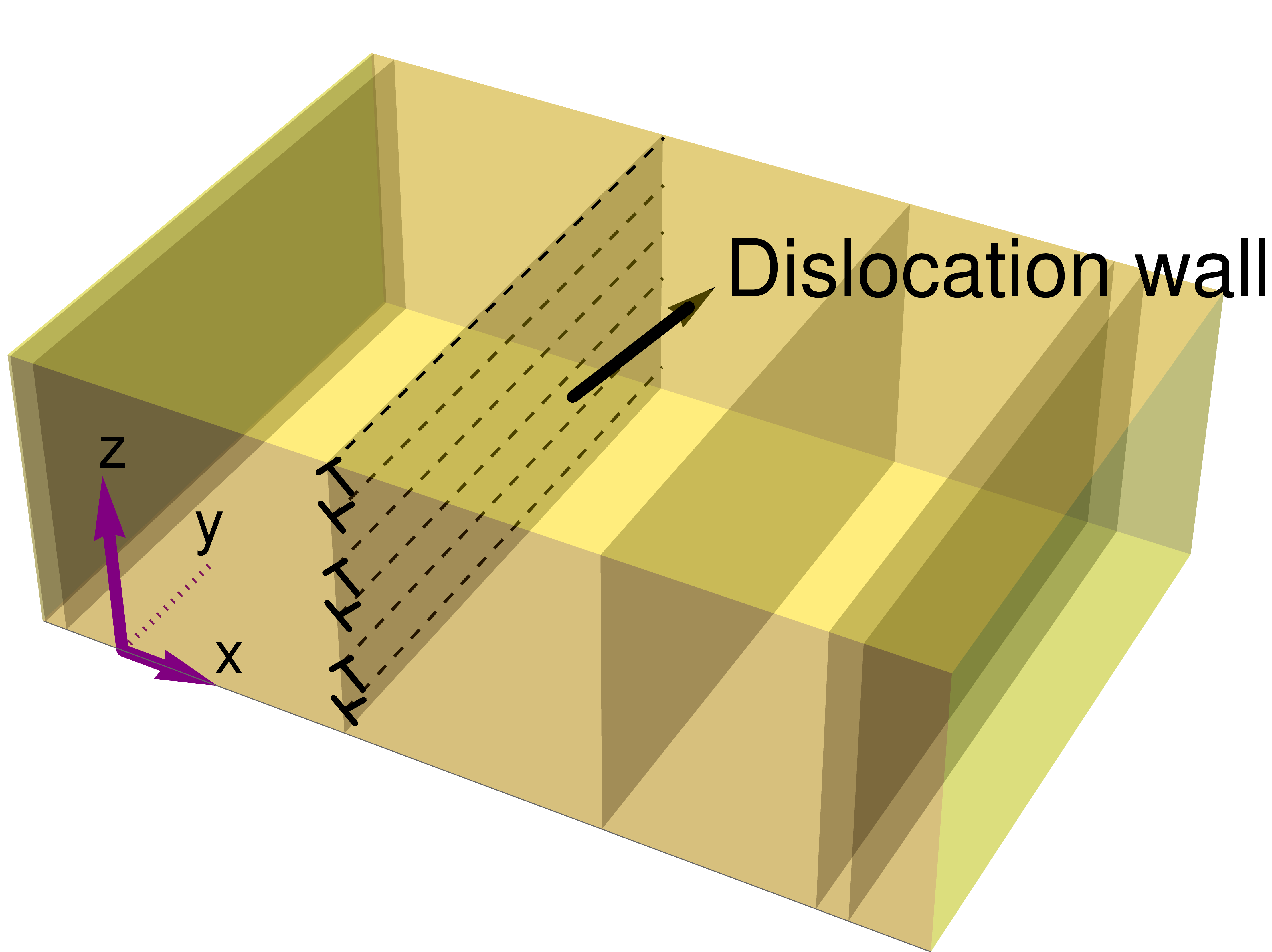}
    \caption{}
\end{subfigure}
\end{minipage}
\begin{minipage}{0.7\hsize}
\begin{minipage}{\hsize}
\begin{subfigure}{0.49\hsize}
\includegraphics[width=\hsize]{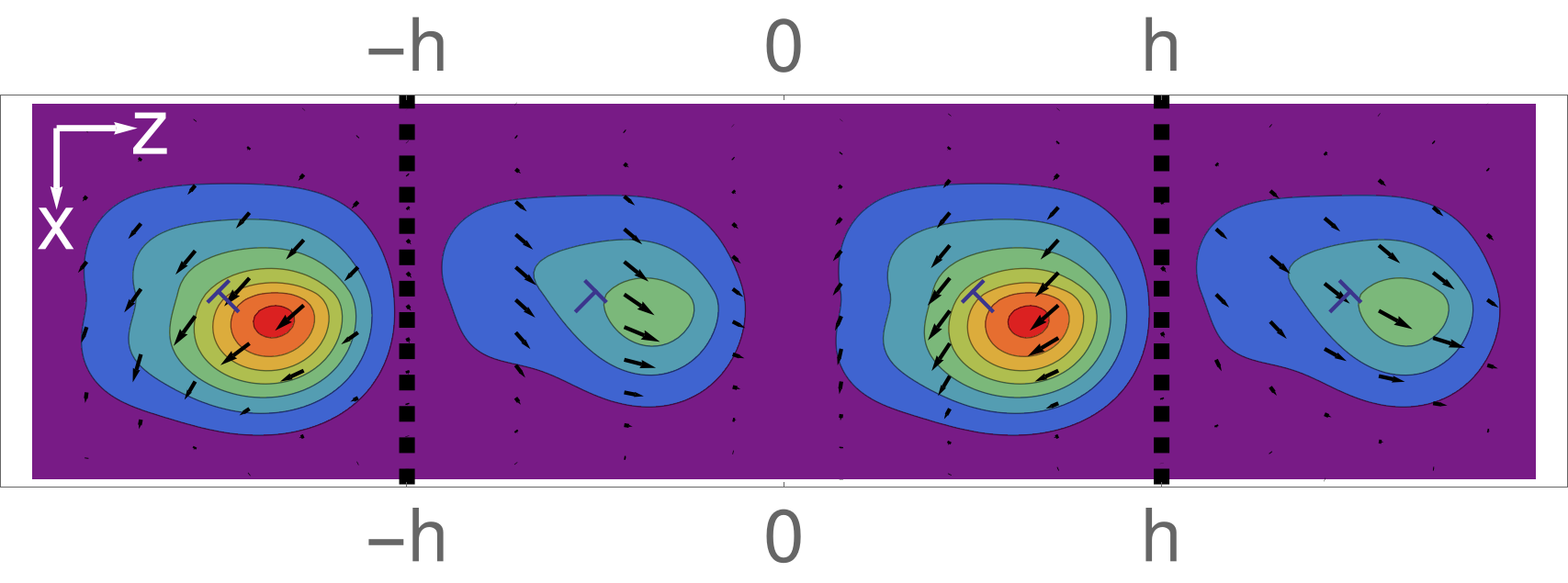}
\caption{}\end{subfigure}
\begin{subfigure}{0.49\hsize}
\includegraphics[width=\hsize,clip,trim={0, -70, 0, -70}]{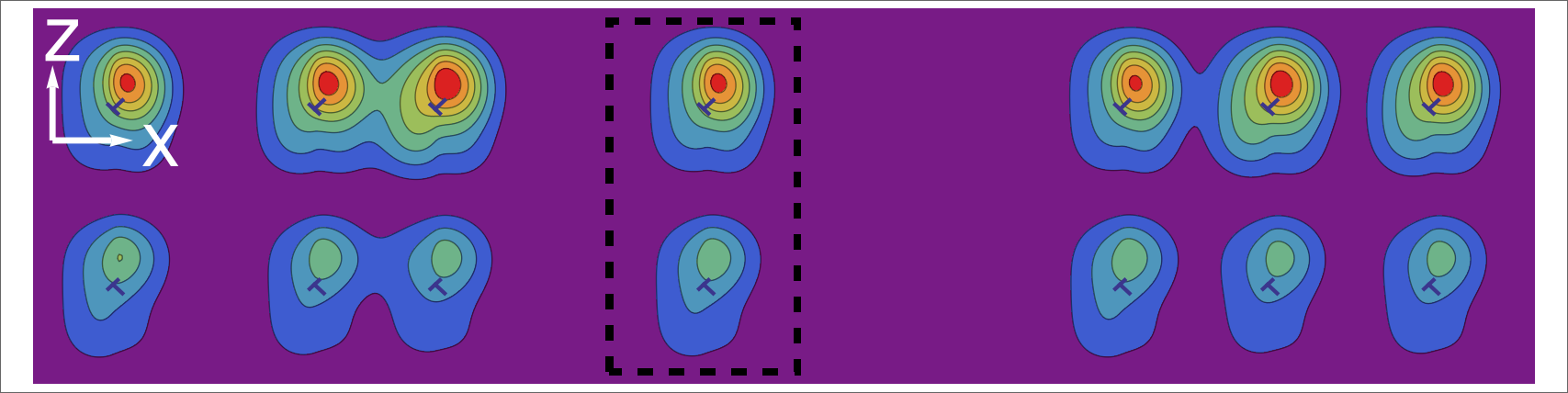}
\caption{}\end{subfigure}
\end{minipage}
\begin{subfigure}{\hsize}
\centering
\includegraphics[width=\hsize,clip,trim={0, 100, 0, 35}]{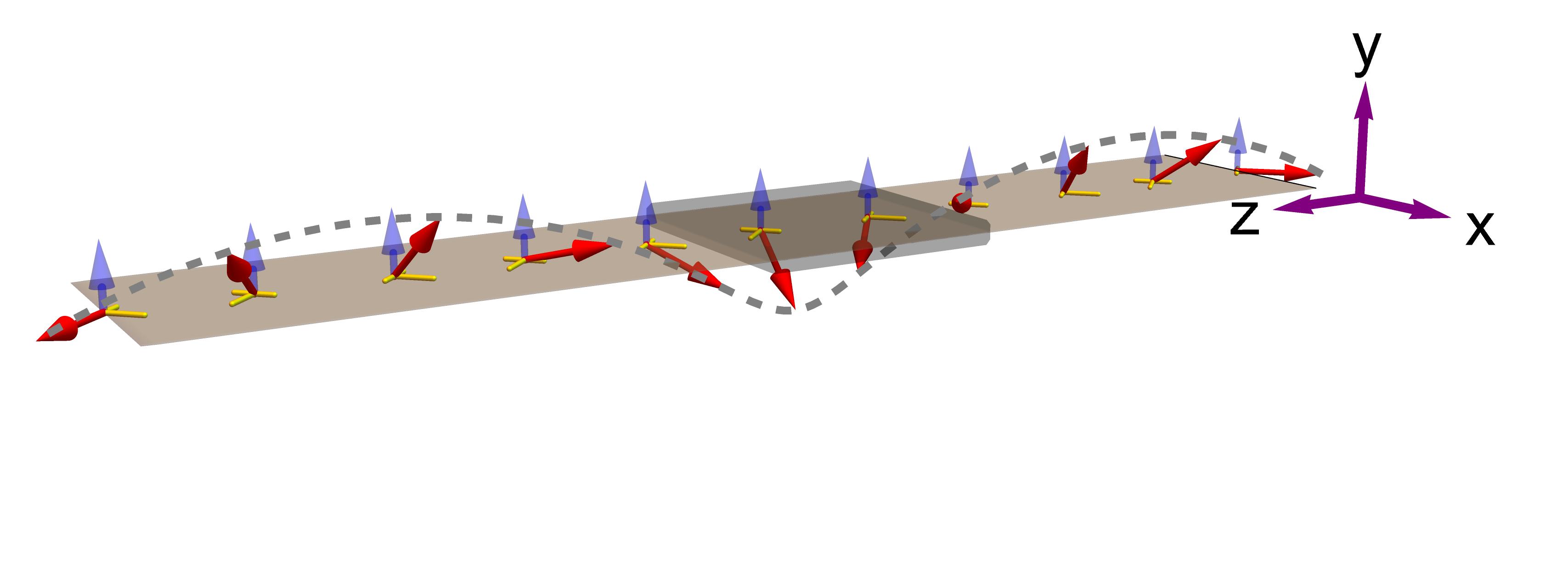}
\caption{}\end{subfigure}
\end{minipage}
\caption{Depiction of the theoretically expected magnetic configuration from several different perspectives. (a) Sketch of the randomly distributed dislocation walls and the Burgers vector orientations \cite{note1}. (b) Magnetic order along a dislocation wall. The dislocation wall consists of a repeating unit cell (dashed lines) of two dislocations (purple Burgers vectors), and elastomagnetic coupling induces a nonzero magnetic moment $\bv{m}$ whose magnitude is strongly localized in ``puddles'' near the dislocations (heatmap, arbitrary units). The planar projection ($m_x,m_z$) of $\bv{m}$ (black arrows) is  shown in an ``antiferromagnetic'' configuration of $m_z$, and ``ferromagnetic'' configuration of $m_x$. This order is nearly degenerate with a fully ferromagnetic one. (c) Magnetic order in a region occupied by a random series of dislocation walls. The dashed region marks the area in the unit cell of panel b. (d) Illustration of a possible spin-spiral structure of the coarse-grained spin-chain model of the dislocation wall. Blue arrows denote the constant background FE field direction, and red arrows the rotating magnetic moments. Only $m_x$ remains ferromagnetic, which explains why there is no out-of-contact magnetic signal.
  % with largest magnitude near the dislocation walls.)
}
  \label{fig:sim-12}
\end{figure*}

%\IS{(Is this paragraph supposed to be actually in the discussion section?)}
%Let us discuss the phenomenology of the magnetism that we found.
The dramatic difference between plastically deformed and pristine samples indicates that lattice degrees of freedom play a crucial role. It is unlikely that the magnetic signal is induced by the tip contact itself, as previous in-contact measurements on undeformed STO and LaAlO${_3}$/STO and $\gamma-$Al$_2$O$_3$/STO interfaces found magnetic signals orders of magnitude smaller \cite{Christensen2019}.
The SQUID signals therefore must arise from deformation-induced regions of both static magnetic order and paramagnetism in close proximity to each other. This is consistent with magnetism localized near the dislocation walls, surrounded by paramagnetic regions.
Furthermore, the magnetic moments must be ordered so that they can only be detected when stress is applied by the tip. This indicates the magnetic order is either homogeneous, i.e. ferromagnetic, but in the plane parallel to the surface ($xy$ plane in Fig. \ref{exp_illustration}a), or an inhomogeneous, e.g., spiral, order. In both these cases the out of plane magnetic field picked up by the SQUID would average to zero, and would only be detected if the tip stress rotated the magnetic order to induce an out of plane ferromagnetic component.
In addition, it is unlikely that the magnetism is triggered by local charge doping in the walls, as the magnetic signal is qualitatively the same for insulating and conducting samples; as noted, the latter are probably superconducting along the same dislocation walls.
We now show that our results are naturally explained by coincident magnetic and FE order, induced directly by the strong deformation of the lattice at the dislocations and the large associated strains, and mutually interacting.

To this end,
we construct an effective low-energy model that describes the interplay between strain, magnetism and FE order. 
For simplicity, we assume that the walls are randomly distributed perpendicular to the deformation, which we choose as the $\hat x$ axis -- see Fig. \ref{fig:sim-12}a for a schematic \cite{note1} . 
Since the dislocation spacing $h \gg a_0$, the physics should be at least qualitatively captured by a continuum theory, and we therefore construct a
Ginzburg-Landau free energy for the coupled magnetic moment $\bv{m}$, strain $\ke_{ij}$, and FE order $\bv{u}$. In STO, $\bv{u}$ in both insulating and conducting samples represents a zone-center polar phonon, whose condensation mediates the breaking of inversion symmetry. Due to Coulomb forces, only the transverse components of $\bv{u}$ are relevant at low temperatures, so we assume $\nabla \cdot \bv{u}\approx 0$.
We neglect additional degrees of freedom, e.g., $\bv{u}$'s corresponding longitudinal mode, or additional zone-boundary phonons that can be expected to only quantitatively change the results. 
We assume that the dislocations in each wall (see Fig. \ref{fig:sim-12}a) run parallel to the scanning surface, since it can be checked that the magnetic signal from dislocations running parallel to the surface will be stronger than those from dislocations terminating at the surface.
Thus, the dislocations have an axis that runs along $\hat y$, and so our fields are (except for negligible boundary effects) independent of $y$, e.g. $\bv{m}=\bv{m}(x,z)$.

We summarize the results of our analysis, details of which are provided in the Methods section. Both $\bv{u}$ and $\bv{m}$ orders are described by simple quadratic free energies, coupled to strain,
\begin{equation}
  \label{eq:free-energy}
  F_\eta = \int dx dz \left[\frac{a_\eta}{2} |\bv{\eta}|^2 + \frac{b_\eta}{4} \sum_j \eta_j^4 + \bv{\eta}\cdot \left(\hat{\Lambda}_\eta\hat{\ke}\right)\cdot\bv{\eta}+\cdots\right],
\end{equation}
where $\eta=m,u$ and $\hat{\Lambda}_\eta \hat{\ke}\equiv \lambda_\eta^{ijkl}\ke_{kl}$ is a strain-coupling tensor whose elements are dictated by the undeformed lattice symmetries.
% Before going into details, we summarize the qualitative features and results of the theory. Strong
The strong strains near dislocations drive both magnetic and FE order,
% strains, such as appear near a dislocation, should drive both magnetic and FE order. The reason is that
since $\ke_{ij}$ couples to quadratic terms of both $\bv{m}$ and $\bv{u}$ in the free energy, and the dipolar strain fields of a dislocation ensure that some regions around the dislocation are always favorable to order.
%the sign of these free energy terms always has attractive regions. 
By construction, the FE order is ``soft'', with $a_u \ll a_m$, consistent with the proximity of clean STO to a FE state, and has strong elastic coupling $|a_u| \ll |\lambda_u|$ \cite{Dunnet2018,Franklin2020}. Hence, the region of induced $\bv{u}$ order is
% Clearly, the FE order, which is \IS{(need to define soft)} softer, will be
far more extensive than the ordered $\bv{m}$ regions. This gives rise to
% . Thus, we may expect
localized ``puddles'' of magnetism around each dislocation, and to quasi-long-range FE order along the walls (see Fig. \ref{fig:sim-12}b,c for a depiction of the calculated magnetic puddles). Interestingly,
% However,
interference effects, the complex unit cell, and the discrete symmetry of even the simplest lattice (we take a cubic lattice), imply complicated enough strain fields to induce
% that the strain fields are very complicated, inducing
nonzero and inhomogenous values in all three components of magnetic field around a given dislocation. In particular, $m_z(x,z) \neq 0$, in contradiction to the absence of a ``no-contact" signal (Fig. \ref{exp_illustration}h). Furthermore, the temperature dependence of $\bv{m}$ (Fig. \ref{exp_illustration}) is also inconsistent with pure strain-induced magnetism,
% if magnetism is interacting only with strain fields, it should have either a
which should lead to Curie-like or, even more probably, a constant-$T$ dependence. 
% To explain the experimental results, w
We must therefore take a step further and investigate the long-range effective interactions among magnetic puddles.

We find, that due to a single 
% Importantly, due to a single
remaining intra-unit-cell $z$-axis inversion symmetry and the relatively large distance between dislocations, different magnetic puddles are approximately magnetically uncoupled, so that a dislocation wall can be thought of as a series of noninteracting spins (one spin per puddle). Thus, the long-range FE order is a natural candidate to mediate long-range spin-spin interactions.
As opposed to the magnetic order, the FE polarization is strongly constrained by its transverse nature. The complicated strain fields always induce longitudinal deformations, except for the case where $\bv{u} = u_y(x,z)\hat{y}$, i.e., when the FE order is parallel to the dislocations themselves.
% This raises the possibility that interactions of these spins with the polar order, which is \IS{(need to define soft)} soft and strongly temperature dependent, are responsible for the long-range order of magnetic moments and its temperature dependence. From symmetry, two types of magneto-polar interactions are expected:
Symmetry constraints imply two allowed couplings: biquadratic ($\sim \bv{m}^2\bv{u}^2$) and magneto-flexupolar, Dzyaloshinskii-Moriya-like ($\sim \bv{u}\bv{m}\nabla\bv{m}$) coupling. The biquadratic term readily accounts for the magnetic $T$ dependence, if the coupling is repulsive.
% readily explains the inverted temperature dependence of the magnetic component if FE and magnetic orders \emph{compete}.
The second term, once the directional constraint of $\bv{u}$ is accounted for,
% will
naturally induces a spiral or helical order for $\bv{m}$ in the $yz$ plane (see Fig. \ref{fig:sim-12}d).
% One further aspect completes the picture: as discussed above, $\bv{u}$ is transverse, since longitudinal polar deformations induce energetically expensive bound charge. However, as we mentioned, the complicated spatial strain fields will always induce longitudinal deformations, unless the FE polarization is strictly parallel to the dislocation axes, $\bv{u} = u_z(x,y)\hat{z}$, running along the $y$-axis walls (see SI for a figure). This insures that the flexupolar coupling prefers a helical configuration with a ferromagnetic moment in the $x$ direction, and a spiral in $y,z$ -- see Fig. \ref{fig:sim-12}d for a schematic.
Such a configuration shows no out-of-plane moment. In the presence of a force provided by the tip,
% The force provided by the tip then creates a z-dependent strain field, again inducing
a longitudinal FE component is again introduced, causing $\bv{u}$ polarization
% the polar moments
to rotate and drag the $\bv{m}$ moments with them.
% The magnetic moments rotate as well,
This induces an out-of-plane ferromagnetic component and a signal in contact.

\section*{Discussion}

The observed magnetism is a result of plastic
strain engineering. Our analysis reveals that the magnetism is induced at the self-organized dislocation walls. Below we discuss a few possible microscopic effects that might be related to the magnetism reported here. 
We note that STO shows a structural transition to a tetragonal phase at 105 K, 
leading to tetragonal domains at low temperatures,
which could generate magnetism as well. While we cannot completely rule out domain wall effects,
their associated strains fileds are
much weaker than those close to a dislocation, and indeed previous measurements of magnetic signals at the walls are two orders of magnitude weaker \cite{Christensen2019}.
%, and while we . Yet we cannot completely rule out domain https://www.overleaf.com/project/636d43a41fb5671594594a4cwall effects, and tailored experiments are needed to clarify their role.

Our results raise two interrelated questions regarding the microscopics of the magnetism: what is the source of the paramagnetism near the dislocations, and what determines the strength of the magnetoelastic coupling that induces static order. Previous studies have shown that STO dislocation strain can change the Ti valence by as much as $-1/2$ \cite{Gao2018}, which raises the possibility that Ti$^{3+}$ states near the dislocations give rise to paramagnetism, analogous to the magnetic properties found in rare-earth titanates such as (Y,La)TiO$_3$ \cite{PhysRevB.104.024410}. Such effects could be the cause of both the paramagnetism and the magnetoelastic properties. Another idea for paramagnetism in STO is that coherent virtual fluctuations of the FE order induce paramagnetic fluctuations near the FE quantum critical point via an analog of Faraday's law \cite{Juraschek2017,basini2022,Dunnet2019}. This mechanism also naturally accounts for the repulsive biquadratic term and the temperature dependence of the magnetism, since the effect is suppressed in the ordered state. However, the magnitude of the magnetic signals that we observe is considerably larger than predicted for this mechanism. It should be possible to distinguish between these two effects at high temperatures in future experiments, where coherent FE motion would be suppressed, but the Ti-ion effects would remain.

%\DP{Detection by other techniques? (probably quite hard for spiral order with a small volume fraction)}
The scanning SQUID technique is particularly suitable for the detection of deformation-induced local magnetism, due to its high spatial resolution and magnetic sensitivity, and the ability to apply force with the tip. Unfortunately, its
% Yet the rather
stringent temperature limitations preclude any investigation of the magnetism at higher temperatures.
% , which is crucial for possible applications.
The use of complementary techniques would therefore be highly desirable, although the conjectured spiral order and small volume fraction (several percent, see Methods)
% nature of the effect presents
present unique challenges to bulk probes such as neutron diffraction, magnetometry, muon spin resonance and NMR.
% Spiral order is difficult to detect with bulk magnetometry, especially if the volume fraction is small; the latter is also an issue for other probes that are sensitive to magnetism, such as NMR and neutron scattering.
Highly sensitive resonant x-ray techniques such as nanobeam magnetic circular dichroism (XMCD) might be the most suitable to shed further light on this effect,
including the intriguing possibility of near-ambient temperature magnetism in deformed STO. 

Conceptually, the most important result of our work is that the large local strains that accompany dislocation structures can induce and enhance quantum materials properties that are otherwise essentially unavailable. Using STO as a model system, we demonstrate the vast potential of plastic deformation as a quantum materials engineering tool.  Some of the key characteristics of this approach are its simplicity, the ability to use macroscopic bulk samples, and a vast available parameter space -- deformation temperature, stress rate and direction can all be varied in a wide range. We anticipate that this approach will enable the creation of self-organized dislocation networks in many other quantum materials of interest, and that other, more tailored approaches could lead to similar results, with further enhanced controllability and the possibility to create micron-scale devices.
Some avenues worthwhile of further investigation are fusing of surfaces to create low-angle grain boundaries, engineering of large epitaxial lattice mismatch, and nanoindentation, all of which are routinely used in materials science to engineer and manipulate dislocation structures.
% Nanoindentation is routinely used in materials science to create and move dislocations, and this could be a fruitful approach in STO and many other systems, since the material does not need to be macroscopically ductile. Moreover, dislocation walls (i.e. low-angle grain boundaries) can be created in a controlled way by fusing two crystal surfaces at elevated temperatures, which provides a high degree of tunability. Finally, in epitaxial thin films and heterostructures, significant lattice parameter mismatch between different components can lead to periodic dislocation arrays. This is usually avoided in thin film growth, but in light of our results it might prove to be an interesting avenue to manipulate electronic properties on the nanoscale.
Overall, we believe that the present work points the way to an exciting possibility of a large-scale quantum-materials engineering effort.

\acknowledgements

We thank N. Bachar, B. Davidovitch, J. Ruhman, P. Volkov, R. M. Fernandes, D. M. Juraschek and A. Balatsky for interesting discussions. We thank A. V. Bj{\o}rlig for performing scanning SQUID measurement on the deformed SrTi$_{1-x}$Nb${_x}$O${_3}$ sample ($x=0.002$). X.W. and B.K. were supported by the European Research Council Grant No. ERC-2019- OG-866236, the Israeli Science Foundation, grant No. ISF-228/22, COST Action CA21144, and the Pazy Research Foundation grant no. 107-2018. A. Kundu and A. Klein acknowledge funding from the Israel Science Foundation (ISF) and the Israeli Directorate for Defense Research and Development (DDR\&D) under grant No. 3467/21. The work at University of Minnesota was funded by the Department of Energy through the University of Minnesota Center for Quantum Materials, under Grant No.
DE-SC0016371. D.P. acknowledges support from the Croatian Science Foundation under grant No. UIP-2020-02-9494. The work at the University of Connecticut was supported by the National Science Foundation award No. 2233149. 

\section{Methods}

\subsection{Samples}

We examined three types of plastically deformed SrTiO${_3}$ crystals: (i) doped SrTi$_{1-x}$Nb${_x}$O${_3}$ ($x=0.002$, Supplementary Information Fig. 3), previously reported to exhibit enhanced superconductivity at low temperatures, prepared by the Minnesota and Zagreb groups from commercial crystals (MTI Corp.) \cite{Hameed2022}, and SrTi${_{1-x}}$Nb${_x}$O${_3}$ ($x=0.004$, Fig. \ref{exp_illustration} and Supplementary Information Fig. 1), sourced from Furuuchi Chemical \cite{Herrera_2019}, plastically deformed by the Connecticut group; (ii) a conducting oxygen-vacancy-doped sample (carrier density $\sim 4 \times 10^{17} \text{cm}^{-3}$, Supplementary Information Fig. 2) and (iii) an undoped, insulating sample (Figs. \ref{exp_illustration} and  \ref{fig:exp_dependence}), both prepared by the Minnesota and Zagreb groups from commercial crystals (MTI Corp.). 

\subsection{Micro-SQUID setup}

The SQUID sensor consisting of a 1.5 $\mu m$ or 7 $\mu m$ diameter pick-up loop was used in experiments at Bar-Ilan or Connecticut, respectively. The pick-up loop detects the magnetic field. Additionally, the sensors have an integrated micro-field coil that applies an alternating local excitation field. 
By combining a low-pass filter and a lock-in technique, we simultaneously record both spontaneous and field-generated magnetism and reliably distinguish between the two.
During the scanning process, the SQUID tip is 0.2-2 $\mu m$ above the sample surface or in contact with it. 
To establish contact, we use piezoelectric elements to press a nonconducting silicon chip onto the sample surface. We record the change of capacitance resulting from the bending of a flexible cantilever during scanning, which corresponds to the force applied by the piezoelectric elements. The force is estimated to be around 0.1 - 1 $\mu$N \cite{Christensen2019}.

\subsection{Estimating the magnitude of the magnetic moments}

Our experiment does not give us a quantitative measurement of the magnetic moments in the sample. However, it is possible to make an order-of-magnitude estimate. Our pickup loop resides at a height $z_0 \sim 1 \mu$m  above the sample surface, and its radius is also $R \sim 1 \mu$m$\sim z_0$. In addition, the contact tip is situated several microns
%about a micron \BK{more like 6 um} ($\sim z_0$) 
from the loop center, and has a submicron
%also has a micrometer scale ($\sim z_0$) 
contact region.
%\BK{i thought it was more like 0.1 um, but the effect extends over 1um}. 
Thus, all relevant lengths are of the same order and for simplicity in what follows we take them all $\sim z_0$. We verified that this approximation is an \emph{under}estimation.

We start with an estimate of the effective magnetization at the surface  disregarding any microscopic considerations such as the dislocation walls. Since strain decays algebraically, we may assume that only magnetic moments within a distance on the order of microns from the tip are rotated out of the plane. We therefore assume a patch of  $\hat z$ polarized surface with a mean surface magnetization $m_S$ and area $\sim z_0^2$. The standard expressions for the magnetic fields yield an average $B \sim (\mu_0/4\pi) m_S / z_0$, where $m_S$ has units of Ampere (magnetization per area). This implies the average flux measured by the SQUID is
\begin{equation}
  \label{eq:squid-flux}
  \Phi \sim \frac{\mu_0 m_S}{4\pi}\left(\frac{R^2}{z_0}\right) \sim \mu_0 m_S z_0.
\end{equation}
% :)
Our mean measured flux at the highest stress is $\Phi \sim 0.01 \Phi_0$, where $\Phi_0$ is the flux quantum. Plugging these expressions back into Eq. \eqref{eq:squid-flux} we find that $m_S \sim \mu_B/a^2$, with a coefficient of order one.

Let us now try to assume that the magnetization is supplied by the dislocation walls, and estimate the magnetization per dislocation. We can obtain a rough estimate of the dislocation density by assuming each dislocation wall removed one unit cell length from the sample, yielding an average inter-wall separation of $d \sim a /\ke$, where $\ke \sim 0.01$ is the deformation of the sample. We assume that the surface magnetization calculated above is supplied by the infinite set of dislocations running parallel to the surface at increasing distances and that the strain rotates dislocations only up to a depth of order $z_0$. In that case, we may estimate $m \sim m_S / z_0$, where $m$ is now the bulk magnetization (in Ampere/meter) from the dislocation distribution. A unit-cell of the dislocation wall has a width of $d$, depth of $h$, and height of $a$. Thus,
\begin{equation}
  \label{eq:magnetization-dis}
  m \sim \frac{m_S}{z_0}\sim \frac{\mu_B}{a d h}\left(\frac{h }{\epsilon z_0}\right) = 0.6 \frac{\mu_B}{a d h}
\end{equation}
so that we have approximately one Bohr-magneton per layer in each dislocation.  We note our estimates are rather conservative since we assumed the tip fully polarizes the magnetic moments and ignored any possible spiral order.

\subsection{Theoretical modeling}

\begin{figure}
\includegraphics[width=0.7\hsize]{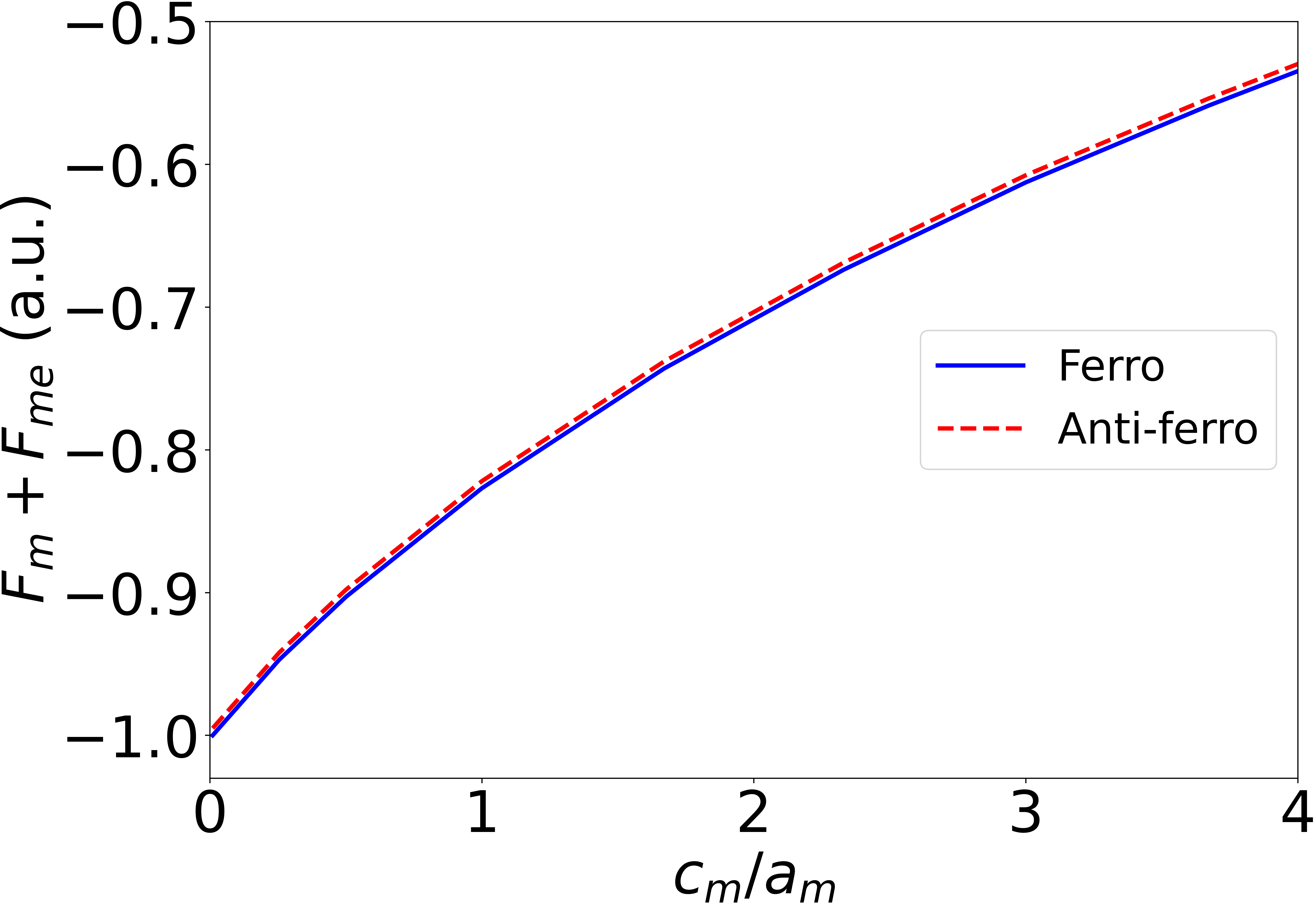}
    \caption{Almost degeneracy of the free energy for ferromagnetic and antiferromagnetic puddle configurations.\label{fig:freeE}}
\end{figure}

%\section{\label{sec:level1} Model}
As described in the main text, we describe our system via a Ginzburg-Landau free energy that couples magnetism, ferroelectricity, and strain. Its form is 
\begin{equation}
\label{eq:F-sum}
    F = F_m + F_u + F_{me} + F_{ue} + F_{mu},
\end{equation}
where $F_m$ denotes the magnetic free energy, $F_u$ denotes the FE free energy, and $F_{me}, F_{ue}$ denote the magneto- and ferro- elastic coupling.
\paragraph*{Magnetic order --}
For the magnetic free energy terms we take,
%The strain-tensor effectively couples such modes and modify the Ginzburg-Landau free energy in following ways, 
\begin{flalign}\label{eq:fmfp}
 F_{m} = \int dx dz \left[\frac{1}{2} a_m |\bv{m}|^2 + \frac{1}{4} b_m\sum_i m_i^4 -\frac{1}{2} c_m \bv{m}\cdot \nabla^2 \bv{m}\right] \\
F_{me}  =\frac{1}{2}\int dx dz
           \left( \lambda^{me}_1\ke_0 |\bv{m}|^2 + \lambda^{me}_2\left[(2\ke_{xx}-\ke_{yy})m_x^2 \right.\right.\nonumber\\
           \left.\left.+(2\ke_{yy}-\ke_{xx})m_y^2 -\lambda^{me}_2\ke_0m_z^2\right)\right],
\end{flalign}
where, $i\in \{1,2,3\}$, $\ke_0=\ke_{xx}+\ke_{yy}$ is the compressive strain, and $\lambda^{mu}_j$ denote respectively coupling to volume-changing, volume-preserving, and shear strain, which are the allowed terms in a cubic lattice (which we take for simplicity). The quadratic-linear coupling implies that $|\bv{m}| \sim \sqrt{\ke}$ upon ordering. We do not know a-priory the values of the various couplings, however, our results do not strongly depend on the values, provided $a_m \lesssim \lambda$. In our numerics we take $\lambda^{me}_3 = 0$ for simplicity.
%%% this  non-analyticity is enough to insure a complicated spatial structure as discussed above.

As discussed above, previous work has established that the dislocation Burgers vectors are positioned approximately equidistantly with a spacing of $h \approx 14 a_0$ in walls lying in the $yz$ plane. We assume that the signal is dominated by dislocations parallel to the measurement $xy$ plane, hence the Burgers vectors in our wall have following pattern (see Fig. \ref{fig:sim-12}), $ \mathbf{b}_n =  b_0\left[\hat{x} +  (-1)^{n} \hat{z}\right];~ \forall~ n \in (-\infty, \infty)$. Each dislocation $\bv{b}_n$ creates a displacement $\bv{w}_n$, with
\begin{align}
\label{displacement}
\bv{w}^{n} & =\frac{(-1)^n b_{0}}{16\pi}\left[\frac{2(x-n z)}{(1-\sigma)r^{2}}\mathbf{b}_n+\frac{(1-2\sigma)\log\left(r^{2}\right)}{(1-\sigma)}\hat{b}_n \right.\nonumber\\
&\left.+4\tan^{-1}\left(\frac{x+n z}{x-n z}\right)\hat{b}_n\right].
\end{align}
where, $r^2=x^2+z^2$ and $\sigma=0.24$ is the Poisson ratio. The strain tensors are as usual $\epsilon_{ij}(x,z)=(1/2)\left(\partial w_i/\partial r_j + \partial w_j/\partial r_i\right)$, and the total strain field is obtained by summing up the infinite series for the wall. (We note that the physical wall is actually half-infinite only. However, this only gives rise to a boundary region of width $\sim h$ which we neglect.) The resulting strain field for a consecutive pair of Burgers vectors oscillates along the $z$  direction, i.e. away from the measurement surface, with periodicity $2h$, and it decays exponentially away from the wall ($\epsilon_{ij}\propto e^{-|x|/h}$, $e^{-2|x|/h}$ at large $|x|$).  We plug the fields back into the free energy, Eq. \eqref{eq:fmfp} and solve the resulting system numerically via a standard relaxation method. Starting from an initially weak spin-ordering this modified coefficient leads to the ferromagnetic/anti-ferromagnetic `puddle'-like structure. For $c_m = 0$, the different puddles are completely independent, and this degeneracy is only weakly lifted by the gradient term, due to a z-axis inversion symmetry (see Fig. \ref{fig:freeE}. As discussed in the main text, for generic coefficients all $m_i \neq 0$.
%%% IAMHERE
%We coarse-grain each puddle chain into a series of effective spins by integrating every half-unit cell, with the resulting moments running along $z$ with magnitude $S_n \sim \left(\lambda_{ME}\ke\right)^{1/2} h^2$.

\begin{figure}
    \includegraphics[width=0.9\hsize]{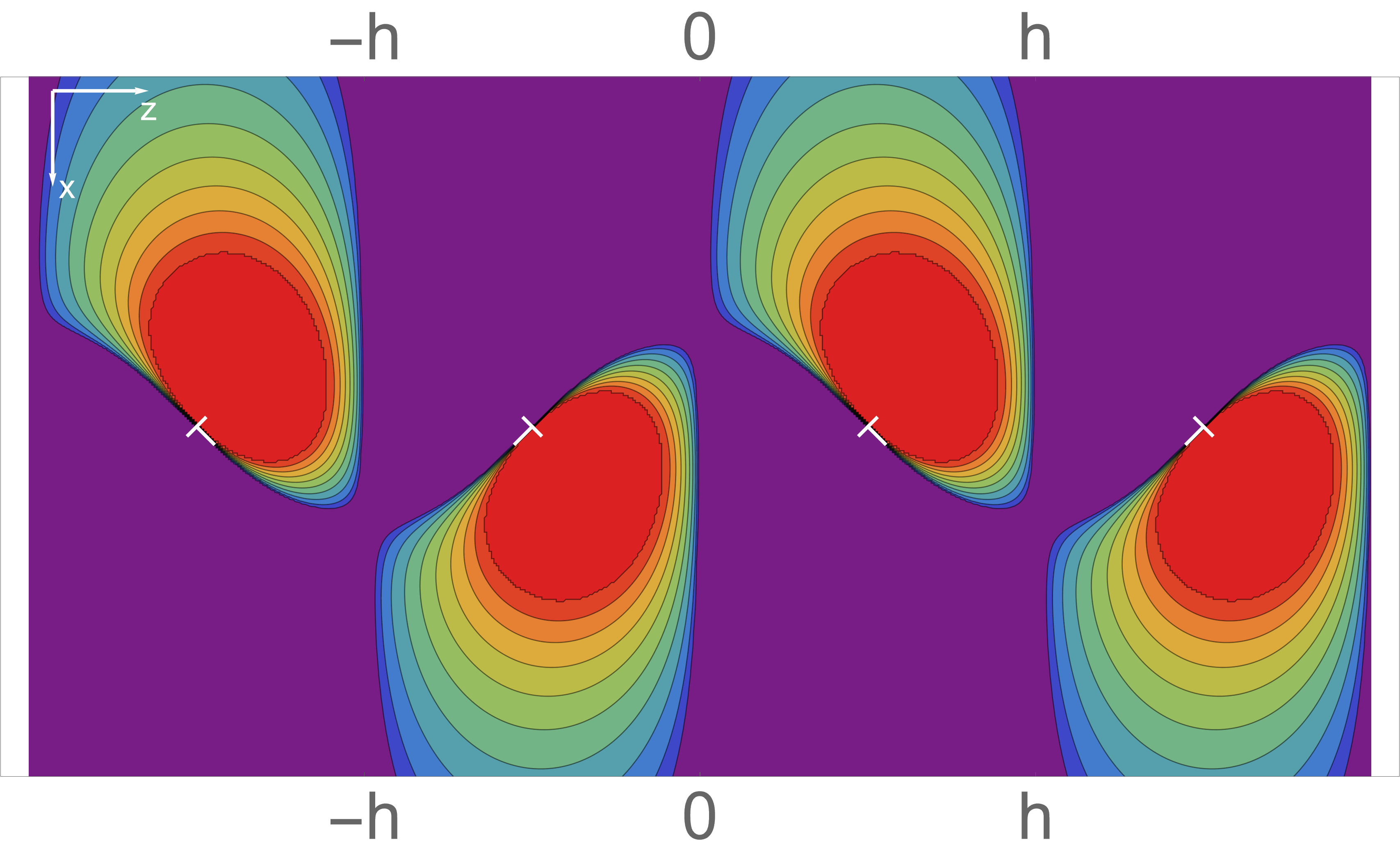}
    \caption{The FE order along the dislocation walls. The intensity map corresponds to $u_y$ (arbitrary units). For simplicity we set $c_u = 0$. At nonzero $c_u$ the ordered regions will merge together and create long-range order.\label{fig:FE}}
\end{figure}
\paragraph*{FE order -- }
The FE free energy has the same form as the magnetic one, see Eqs. \eqref{eq:free-energy} and \eqref{eq:fmfp}. The coefficients $a_u, \lambda_j^{ue}$ have been previously estimated from DFT \cite{Hameed2022,Dunnet2018}, which means that $\bv{u}(x,z)$ can be computed explicitly, at least for $c_u = 0$. Taking $c_u >0$ does not qualitatively change the result. As discussed in the main text, since $\nabla \cdot \bv{u} \approx 0$, the FE order must be parallel to the dislocation lines, i.e. $\bv{u} = u_y(x,z) \hat y$, see Fig. \ref{fig:FE}.
%The polar modes has similar form as magnetic modes. Below $T_c$ the polar modes are soft, which implies that the polar moments will be much less localized than the magnetic moments, which is consistent with previous Raman measurements showing extended symmetry breaking due to dislocation. Besides, the polar term arises only from the transverse phonon mode, meaning $\sum_i\pd_i P_i = 0$, which makes it evident from the GL saddle point equation that the polar modes having only non-zero component along $z$-axis, i.e., $\bv{P} = (0,0,P_z)$. 
 
%{\it Temperature dependence:}
\paragraph*{The magneto-polar coupling --}
Next we consider the possible magneto-polar interactions. The simplest coupling is just fully-symmetric under all lattice transformations, namely, $F_{mu} = \int dx dz\frac{1}{2}Q_{mu} |\mathbf{m}|^2|\mathbf{u}|^2$. Previous experiments in deformed STO found that  the FE mode decreases (roughly linearly) with temperature up to some rather high $T_{FE} > 100$K. Thus, the temperature dependence $\bv{m}(T)$ depicted in Fig. \ref{fig:exp_dependence}f is consistent with $Q_{mu} > 0$. Indeed, the coupling just renormalizes the magnetic $a_m$, namely, $a_m \to a_m + Q_{mu} |\mathbf{u}|^2 \approx a_m + Q_{mu} \frac{A_0 + a_0 (T_{FE} - T)}{b_u}$ where $A_{0}$ is the average strain-induced correction to $a_u = a_0 (T_{FE}-T)$. Crucially, we see that $a_m$ \emph{decreases} and therefore $|\bv m|$ \emph{increases} with increasing temperature. 
The second possible coupling is a magneto-polar-elastic coupling where $\bv{u}$ appears linearly and inversion symmetry is restored by a gradient operator. To understand the impact of this coupling, we recall that
%{\it magneto-polar coupling:} 
a key difference between the magnetic and FE modes is, that the magnetic moments are localised forming `puddles' along the dislocation lines, but the FE mode has long-range order. We therefore coarse-grain each dislocation wall into a series of ``spins'' running along $z$ with magnitude $m_n \sim \left(\lambda^{me}\ke\right)^{1/2} h^2$, and similarly for the FE mode. This means we can treat each wall as a 1D chain, and for simplicity we assume the spins have constant magnitude $m_0$, and the FE order is constant in both magnitude and direction. In that case the lowest order coupling is just
%
%STO is known to have strong flexo-electric properties.
% For our deformed STO, the symmetry allow us the magneto-polar-elastic free-energy which effectively couples the magnetic orders between 'puddles' along dislocation line ($z$-axis). Simplest form of such coupling can be written as, very simple form,
$F_{mue} = \int dz Q_{mue} m_y(z) u_y \pd_z m_z(z) $, with $Q_{mue}$ being the coupling constant. 
%where we neglected some irrelevant biquadratic terms. $\lambda^{MP}>0$ accounts for the temperature dependence of the magnetic signal (competing orders scenario). 
The gradient term induces a helical configuration for the spins along the out-of-plane $z$-direction. To see this it is convenient to study the system in the single-mode approximation $m_x = \mbox{const}, m_y = m_0 \cos(qz), m_z = m_0 \sin(qz)$. A solution exists with a nonzero $q = - Q_{mue} u_y / (2 c_m)$, which is a minimum of the free energy for $u_y^2 > N_{mue} b_m c_m m_0^2 /Q_{mue}^2$,
where $N_{mue} = O(1)$ is some numerical coefficient. Note that when $c_m$ goes to zero the spiral is always preferred.

\bibliographystyle{ieeetr}
% \bibliography{bibtex-plastically-deformed-sto-vr1}
\bibliography{bibtex-stoStrain}

\end{document}

\title{
%Geometrically induced multiferroic properties in plastically deformed SrTiO$_3$ \IS{alternative title 
Extended data for: Multiferroicity in plastically deformed SrTiO$_3$}

\maketitle

\begin{figure*}[h]
\centering
\includegraphics[width=1\textwidth]{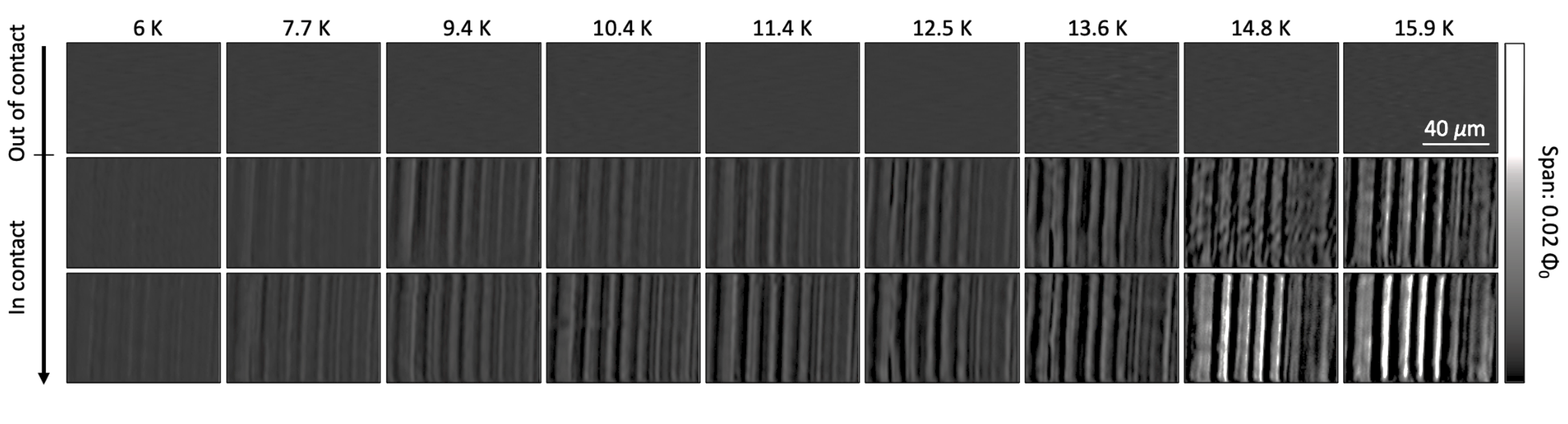}
\caption{Stress dependence and temperature dependence of magnetic signal in plastically deformed conducting sample (SrTi${_{1-x}}$Nb${_x}$O${_3}$, $x=0.004$), shown in Fig. 1f,g. }
\label{SI, Nb-doped 1}
\end{figure*}

\begin{figure*}
\centering
\includegraphics{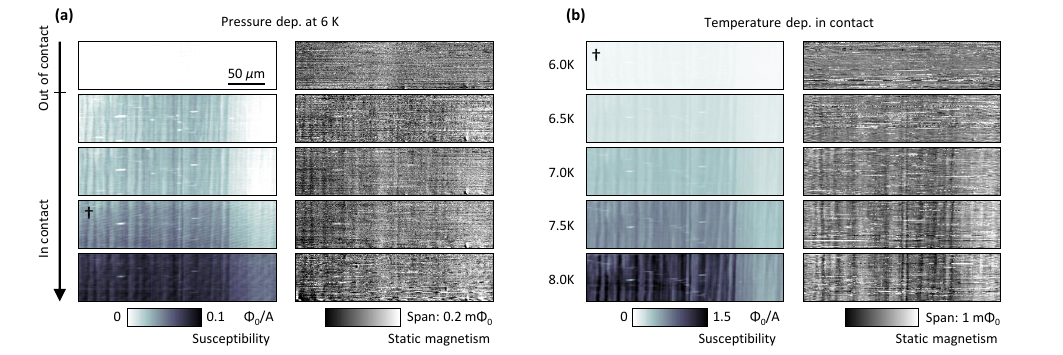}
\caption{Stress and temperature dependence of magnetic signal in a conducting oxygen-vacancy-doped, plastically deformed SrTiO${_3}$. (a) Stress dependence of magnetic stripes at 6 K. (b) Temperature dependence of magnetic stripes in susceptibility and static magnetism by scanning in contact. Images marked with a small dagger are scans performed in the same condition. In this experiment, the contact is nonuniform along the horizontal direction.}
\label{SI, OVD}
\end{figure*}

\begin{figure*}
\centering
\includegraphics[width=1\textwidth]{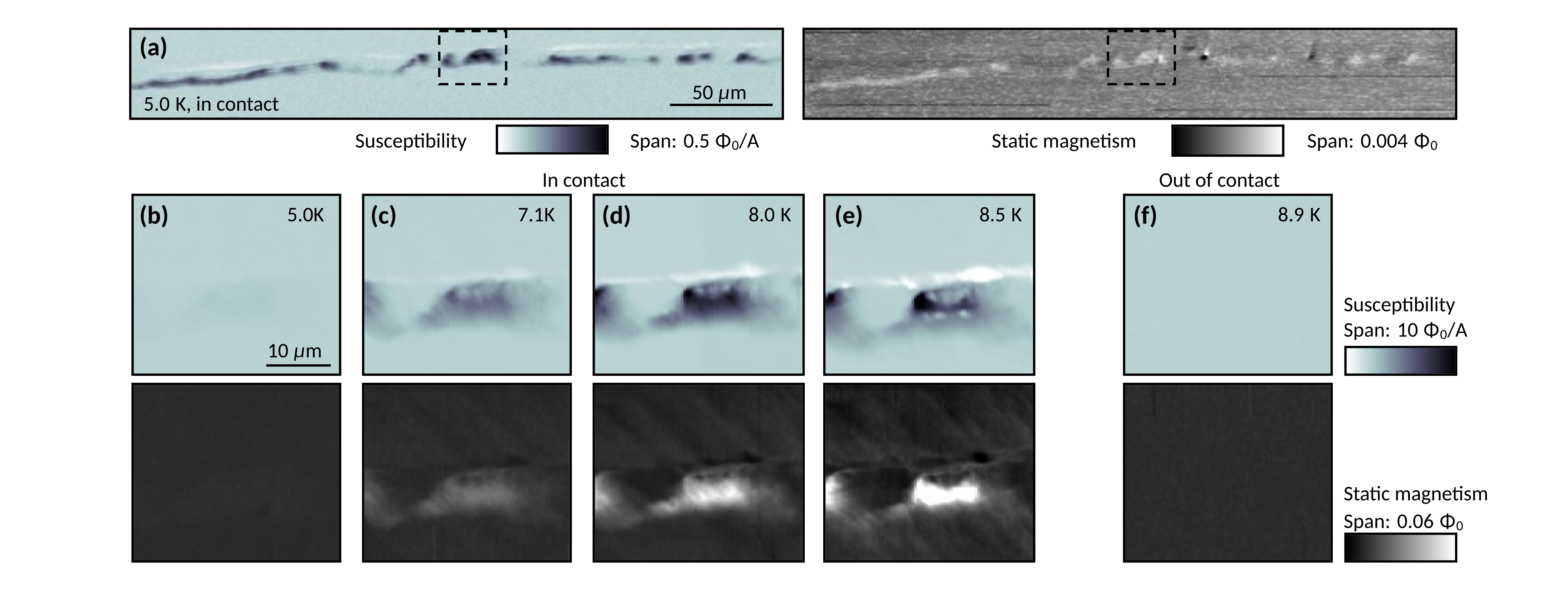}
\caption{Magnetic signal upon contact in a Nb-doped, plastically deformed sample (SrTi${_{1-x}}$Nb${_x}$O${_3}$, $x=0.002$). (a) The signal is only found in a small region on the surface upon contact. (b-e) Temperature dependence of magnetic signal in susceptibility and static magnetism by scanning in contact. The scan region is the dashed rectangle in panel a. The magnetic response increases with the temperature. (f) Magnetic signal disappears when scanning in non-contact mode.}
\label{SI, Nb-doped 2}
\end{figure*}

%\bibliographystyle{unsrt}
% \bibliography{bibtex-plastically-deformed-sto-vr1}
%\bibliography{bibtex-stoStrain}